\documentclass{nature}

\usepackage{lineno}

\usepackage{graphicx}
\makeatletter
\let\saved@includegraphics\includegraphics
\AtBeginDocument{\let\includegraphics\saved@includegraphics}
\renewenvironment*{figure}{\@float{figure}}{\end@float}
\makeatother

\title{An extended hydrogen envelope of the extremely hot giant exoplanet KELT-9b}

\author{Fei Yan$^{1*}$, Thomas Henning$^{1}$} 

\begin{document}

\maketitle

\begin{affiliations}
 \item Max Planck Institute for Astronomy, K\"onigstuhl 17, 69117 Heidelberg, Germany
\end{affiliations}

\begin{abstract}
Giant exoplanets orbiting close to their host stars have high temperatures because of the immense stellar irradiation which they receive. The extreme energy input leads to the expansion of the atmosphere and the escape of neutral hydrogen \cite{Vidal-Madjar2003} \cite{Ballester2007}  \cite{Ehrenreich2015}.
A particularly intriguing case among the hot giant planets is KELT-9b -- an exoplanet orbiting very close to an early A-type star with the highest temperature ($\sim$ 4600 K at day-side) among all the exoplanets known so far \cite{Gaudi2017}. The atmospheric composition and dynamic of such a unique planet have been unknown.
Here we report the first detection of an extended hot hydrogen atmosphere around KELT-9b.
The detection was achieved by measuring the atomic hydrogen absorption during transit with the Balmer $\mathrm{H_\alpha}$ line, which is unusually strong mainly due to the high level of extreme-ultraviolet radiation from the star. 
We detected a significant wavelength shift of the $\mathrm{H_\alpha}$ absorption which is mostly attributed to the planetary orbital motion \cite{Snellen2010}.
The obtained transmission spectrum has a significant line contrast ($1.15\%$ extra absorption at the $\mathrm{H_\alpha}$ line centre). 
The observation implies that the effective radius at the $\mathrm{H_\alpha}$ line centre is $\sim 1.64$ times the size of the planetary radius, indicating the planet has a largely extended hydrogen envelope close to the size of the Roche lobe ( $1.91_{-0.26}^{+0.22}~R_\mathrm{planet}$) and is probably undergoing dramatic atmosphere escape.

\end{abstract}

We observed two transits of KELT-9b on August 6 and September 21, 2017 with the CARMENES instrument \cite{Quirrenbach2016}. CARMENES is a high-resolution (R $\sim$ 94 600 in the visual channel) spectrograph installed at the 3.5 m telescope of the Calar Alto Observatory. Each of the observations lasted for $\sim$ 6 hours covering the 4 hours transit and 1 hour before and after transit. 

For each transit dataset, we firstly removed the telluric $\mathrm{H_2O}$ lines at the $\mathrm{H_\alpha}$ wavelength range and then aligned all the spectra to the stellar rest frame. A reference spectrum was obtained by combining all the spectra observed during out-of-transit. This reference spectrum is regarded as the spectrum of the star and each observed spectrum was then divided by the reference spectrum to remove the stellar  $\mathrm{H_\alpha}$ spectrum. In this way, we obtained planetary absorption spectra with telluric and stellar lines removed. We then binned the planetary absorption spectra from both nights with a 0.01 orbital phase step, i.~e. the spectra within each 0.01 phase bin were averaged with the spectral signal-to-noise ratio (SNR) as weight.

The obtained planetary absorption spectra are displayed in Fig.1a. The figure shows a black shadow during transit which is the result of the $\mathrm{H_\alpha}$ absorption. The absorption feature has a significant wavelength shift and this radial velocity (RV) change is mostly attributed to the planetary orbital motion. 
We modeled the planetary absorption feature assuming it has a Gaussian profile. The semi-amplitude of the planetary orbital motion($K_\mathrm{planet}$) was set as a free parameter in the model. We included an atmosphere wind speed ($v_\mathrm{wind}$) in our model since high-altitude wind from day-side to night-side has been observed in exoplanet transmission spectra, for example in HD189733b \cite{Brogi2016} \cite{Wyttenbach2015} \cite{Louden2015} and in HD209458b \cite{Snellen2010}.

In addition to the planetary $\mathrm{H_\alpha}$ absorption, there is also variation of the stellar line during transit due to the Rossiter-McLaughlin (RM) effect and the Centre-to-Limb Variation (CLV) effect. The RM effect \cite{Rossiter1924} \cite{McLaughlin1924} \cite{Queloz2000} means that the observed stellar line profile changes during transit because of the stellar rotation. The RM effect of KELT-9b has been observed in ref. \cite{Gaudi2017} with the Doppler tomography method and the star is rapidly rotating ($v~\mathrm{sin}I = 111.4$ km/s) with a spin-orbit alignment ($\lambda$) of $-84.8$ degrees (a nearly polar orbit). We adopted these parameters to model the RM induced variation of the stellar $\mathrm{H_\alpha}$ line.
The CLV effect describes the variation of stellar lines across the stellar disk's centre to limb \cite{Yan2015a} \cite{Czesla2015} \cite{Yan2017}. 
We modeled the stellar $\mathrm{H_\alpha}$ line variation considering both RM and CLV effects (see Methods). However, the model is for one planetary radius ($R_\mathrm{planet}$),  the actual effective radius is larger than $R_\mathrm{planet}$ because of the planetary absorption. Thus we introduced a factor $f$ to account for a larger planetary effective radius.

We applied Markov Chain Monte Carlo (MCMC) simulations with the \textit{emcee} tool \cite{Mackey2013} to model the observed spectra and to perform the error estimations (see Methods). 
The best-fit model is presented in Fig.1b. The model consists of two parts: the planetary absorption and the stellar line profile change. The obtained $f$ factor has a value of 1.98.

The semi-amplitude of the planetary orbital motion is derived to be $K_\mathrm{planet} = 268.7_{-6.4}^{+6.2} ~ \mathrm{km/s}$ ( all the errors are 1 $\sigma$ errors). This corresponds to a RV change from -91 km/s to +91 km/s during transit (indicated as blue dashed line in Fig.1b). 
By combining the measured  $K_\mathrm{planet}$ and the orbital motion of the host star  $K_\mathrm{star}$, one derives directly the masses of the star and the planet using Newton's law of gravity. The RV of the host star orbiting the mass centre of the star-planet system has been previously observed to have a semi-amplitude of $K_\mathrm{star} = 0.276\pm0.079 ~ \mathrm{km/s}$. Considering the orbital period of 1.48112 days, orbital inclination of 86.79 degrees and a zero eccentricity, we derived the stellar mass $M_\mathrm{star} = 3.00\pm0.21 ~ M_\mathrm{Sun}$ and planetary mass $M_\mathrm{planet}  = 3.23\pm0.94 ~ M_\mathrm{Jupiter}$. The stellar mass obtained with this method is generally consistent with the mass derived from stellar spectrum modeling ($M_\mathrm{star} = 2.52_{-0.20}^{+0.25} ~ M_\mathrm{Sun}$).

We further obtained the resolved $\mathrm{H_\alpha}$ absorption line profile by adding up all the fully in-transit (excluding the ingress and egress) transmission spectra with the planetary orbital motion RV corrected. We also corrected the stellar induced line profile change due to RM and CLV effects.
The final $\mathrm{H_\alpha}$ absorption line is presented in Fig.2. A Gaussian profile with parameters obtained from the MCMC analysis is also plotted. The absorption line is well resolved with a line contrast of $1.15\pm{0.05}\%$ and a full width half maximum of $51.2_{-2.5}^{+2.7}~\mathrm{km/s}$. 
The line has a Doppler shift of $-1.02_{-1.00}^{+0.99}~\mathrm{km/s}$, indicating we did not detect a significant day- to night-side wind considering the error.

Fig.3 shows the time-series equivalent width of  the $\mathrm{H_\alpha}$ absorption ($W_\mathrm{H_\alpha}$). The equivalent width is obtained by integrating the absorption feature from -75 km/s to +75 km/s ($\sim$ 3.28 $\mathrm{\AA}$) (see Methods). The stellar induced line profile change was corrected before the integration. The integrating band is centered at the planetary RV (i.~e. orbital motion RV plus wind speed). The $W_\mathrm{H_\alpha}$ value shows clear implication of absorption during transit. The $W_\mathrm{H_\alpha}$ quantity for the fully in-transit spectra has an average value of $0.014 \pm 0.002~\mathrm{\AA}$. We did not detect any significant pre-transit or after transit $\mathrm{H_\alpha}$ absorption signal. 

Extended atomic hydrogen atmospheres have been observed for several exoplanets. These observations were mostly performed with $\mathrm{Ly_{\alpha}}$ line using the STIS ultraviolet spectrograph mounted on Hubble Space Telescope  \cite{Ehrenreich2015} \cite{Etangs2012} \cite{Lavie2017}. However, since KELT-9b has a large distance to Earth, it is not possible to detect $\mathrm{Ly_{\alpha}}$ due to the strong interstellar medium absorption.

Detecting the hydrogen atmosphere with the $\mathrm{H_\alpha}$ line is more difficult because it is not a resonant line like the $\mathrm{Ly_{\alpha}}$ and requires a large number of excited atoms in the atmosphere.
There are $\mathrm{H_\alpha}$ detections in one of the most extensively studied exoplanets HD 189733b  \cite{Jensen2012} \cite{Cauley2015} \cite{Cauley2016}, although the detections are affected by stellar activity and vary during different transit epochs \cite{Barnes2016} \cite{Cauley2017}.
Theoretical $\mathrm{H_\alpha}$ modeling work \cite{Christie2013} \cite{Huang2017} shows that extreme-ultraviolet (EUV) and $\mathrm{Ly_\alpha}$ radiation are needed to produce a large number of excited hydrogen atoms, and in the case of HD 189733b, it is probably the stellar activity which provides such a strong EUV radiation considering its host star is a K-type.
As an early A-type star (detailed spectral type: B9.5-A0), KELT-9 (HD 195689) has a strong EUV radiation even without significant activities  \cite{Gaudi2017}.
Thus, the observed strong $\mathrm{H_\alpha}$ absorption in KELT-9b's atmosphere is probably the result of an immense EUV radiation received and an extremely high temperature achieved.
Effects like stellar variability and stellar activity (including enhanced chromospheric activity due to star-planet interaction\cite{Shkolnik2008} ) could potentially affect the $\mathrm{H_\alpha}$ detection.
Although we are not able to measure the activity level of KELT-9 since CARMENES does not cover the Ca II H\&K lines, early A-type stars like KELT-9 normally do not exhibit strong activities because they almost have no chromosphere or corona. The same level of $\mathrm{H_\alpha}$ absorption observed during two transit epochs (see Results of individual night in Methods) also suggests that the host star is probably not very active.
Furthermore, the Doppler information of the $\mathrm{H_\alpha}$ absorption originated from the planet orbital motion makes our detection unambiguous.


The observed extra absorption depth at the $\mathrm{H_\alpha}$ line centre is $1.15 \pm 0.05\%$ while the photometric transit depth in the continuum is only $0.68\%$\cite{Gaudi2017}. This implies an effective radius at the line core of $\sim 1.64 ~ R_\mathrm{planet}$. 
According to theoretical calculations \cite{Huang2017}, $\mathrm{H_\alpha}$ absorption is mostly originated from an atomic layer of the planetary atmosphere and the excited hydrogen atoms at n=2 state have a roughly constant number density ($n_2$) in the atomic layer. 
Therefore, we inferred the observed $\mathrm{H_\alpha}$ absorption in KELT-9b is probably from an atomic layer extending from $1 ~ R_\mathrm{planet}$ to $1.64 ~ R_\mathrm{planet}$ with a temperature range from $\sim$ 4600 K (day-side temperature) to $\sim$ 10000 K.  
The $\mathrm{H_\alpha}$ absorption is optically thick for the sight-line passing the lower part of the atomic layer, and thus the line profile is dominated by the thermal broadening and the maximum line centre optical depth  $\tau_0$. The observed $\mathrm{H_\alpha}$ line profile in KELT-9b has a much larger line width ($21.7 \pm1.1$km/s, $\sigma$ of the Gaussian fit) than the thermal broadening velocity (10.8 km/s, assuming an average temperature of 7000 K), which demonstrates that the $\mathrm{H_\alpha}$ absorption is optically thick.
Using the analytic model in ref.\cite{Huang2017}, we estimated that an excited atom number density of $n_2 = 2.7  \times 10^3 \mathrm{cm}^{-3}$ (which corresponds to a maximum line centre optical depth $\tau_0$ = 57) can produce the observed $\mathrm{H_\alpha}$ line width.
Contributions from rotational broadening as well as possible hydrodynamic escaping broadening \cite{Tian2005} are relatively small and therefore neglected in the above estimation. 

As KELT-9b is orbiting very closely to its massive host star, the planetary Roche lobe has a relatively small size.
The Roche radius at the planetary terminator (where the transit absorption occurs) is smaller than the radius towards the star \cite{Etangs2007}.  Therefore we used the effective Roche radius as an approximation of the value at the terminator \cite{Ehrenreich2010},
and the obtained radius is $R_\mathrm{Roche} = 1.91_{-0.26}^{+0.22}~R_\mathrm{planet}$. 
The observed effective radius at the $\mathrm{H_\alpha}$ line core is close to the Roche lobe, implying that the hydrogen atmosphere almost fills the Roche lobe and can further escape the planet. We estimated the mass loss rate by assuming a hydrogen particle at  $1.64 ~ R_\mathrm{planet}$ can escape the planet if the particle can reach the Roche lobe with a large enough kinetic energy. Using the above calculated density of $n_\mathrm{2}$ and assuming $n_\mathrm{2} / n_\mathrm{H} = 10^{-6}$ at the atomic layer boundary, we obtained a mass loss rate of $\sim 10^{12}~\mathrm{g~s^{-1}}$ for $R_\mathrm{Roche} = 1.91~R_\mathrm{planet}$.
However, since the Roche radius has large errors (1 $\sigma$ range: $1.65 \sim 2.13 R_\mathrm{planet}$) due to the not well determined planetary mass, the mass loss rate estimated from this method has one order of magnitude uncertainty. 
The estimated escape rate here is in broad agreement with the value in the discovery paper \cite{Gaudi2017} ($10^{10} - 10^{13} ~\mathrm{g~s^{-1}}$) which is estimated using the EUV flux level.
Note here we only considered thermal escaping to obtain a rough estimation, the actual escape rate calculation requires a comprehensive modeling work (e.~g. ref.\cite{Salz2016}).



Given the high temperature of KELT-9b ($\sim$ 4600 K at day-side \cite{Gaudi2017}) and an extremely high level of ultraviolet radiation received, most of the molecular hydrogen is probably dissociated into atomic hydrogen and further been stripped away. 
Further transit measurements of other Balmer lines (e.~g. $\mathrm{H_\beta}$ ) will enable a comprehensive study of the planetary temperature structure, and the combination of the Balmer line modeling and the atmosphere escape modeling will allow to constrain the mass loss rate as well as the atmosphere life time more precisely.

\begin{addendum}
 \item[Correspondence] Correspondence and requests for materials
should be addressed to Fei Yan (email: fyan@mpia.de).
 \item[Acknowledgements] This work is based on observations collected at the German-Spanish Astronomical Center, Calar Alto, jointly operated by the Max-Planck-Institut f\"ur Astronomie Heidelberg and the Instituto de Astrof\'isica de Andaluc\'ia (CSIC). We thank Calar Alto Observatory for allocation of director's discretionary time to this programme. 
We thank Lisa Nortmann for help on observing preparations and Matthias Samland for discussions on Monte Carlo analysis. We thank Karan Molaverdikhani and Chenliang Huang for discussions on atmosphere model.
 \item[Author contributions] F. Y. planed the observations, performed the data analysis and wrote the manuscript. Th. H. contributed to observing preparations and writing the manuscript.
 \item[Competing Interests] The authors declare that they have no competing financial interests.
\end{addendum}

\pagebreak


\pagebreak 

\begin{methods}

\subsection{Observations.}
We observed two transits of KELT-9b with CARMENES, a fiber-fed echelle spectrograph mounted on the 3.5 meter telescope at Calar Alto. The instrument has two arms: one in the near infrared and one in the visible. We used only the data taken with the visible spectrograph which has a resolution of R = 94 600 and a wavelength coverage from 0.52 $\mathrm{\mu m}$ to 0.97 $\mathrm{\mu m}$ with 42 spectral orders.
The host star is a bright early A type star (magnitude V=7.56 mag, $T_\mathrm{eff}$ = 10,170 K). The exposure times were set to 300 or 400 seconds. The observation lasted for $\sim$ 6 hours for each night.
Both of the two observations were performed under partially cloudy weather condition, thus some of the spectra have relatively low signal-to-noise raios and part of the observations was lost. 

\subsection{Data reduction.}
Standard data reduction of the raw spectrum was performed with the pipeline CARACAL (CARMENES Reduction And Calibration\cite{Caballero2016} ), including bias, flat, cosmic ray correction and wavelength calibration. The spectra were further reduced with a self-written IDL script with three main steps. 

\textit{Step 1 Removal of telluric lines.} The telluric features at the $\mathrm{H_\alpha}$ wavelength range are mostly $\mathrm{H_2O}$ lines. The telluric removal was performed in the observer rest frame. We firstly built a theoretical $\mathrm{H_2O}$ transmission model with the HITRAN database\cite{Rothman2009} based on the method described in \cite{Yan2015}. The next procedure was to calculate the $\mathrm{H_2O}$ column densities for each spectrum as the value varied during the observations. The column density was calculated by fitting strong $\mathrm{H_2O}$ absorption lines from an adjacent spectral order with a wavelength range of 6474 $\mathrm{\AA}$ $\sim$ 6497 $\mathrm{\AA}$. The observed spectra were normalized before the fitting.
The column density value was then used to calculate the theoretical $\mathrm{H_2O}$  spectrum at the $\mathrm{H_\alpha}$ wavelength range. The telluric lines were removed by dividing each observed spectrum with the corresponding modeled $\mathrm{H_2O}$ transmission spectrum.

\textit{Step 2 Removal of stellar lines.} The stellar line removal was performed in the stellar rest frame. The RV correction includes three parts: (1) barycentric Earth radial velocity; (2) absolute stellar RV ($-20.567\pm0.1$ km/s)\cite{Gaudi2017}; (3) stellar orbital motion RV around the star-planet mass centre. The stellar orbital RV was obtained with $K_\mathrm{star} = 0.276$ km/s from the literature. After these RV corrections, the stellar photospheric $\mathrm{H_\alpha}$ lines were well aligned in all the spectra. We then built a reference spectrum ($F_v^\mathrm{out}$) by combing all the out-of-transit spectra with the SNR as weight. Each of the observed spectrum was then divided by this reference spectrum to remove the stellar  $\mathrm{H_\alpha}$ line. In this way, we obtained the residual spectrum at different phases $R_v=F_v/F_v^\mathrm{out}$. Here $v$ is wavelength expressed in velocity space ($v$ = 0 corresponding to the stellar line centre at 6562.81 $\mathrm{\AA}$)

\textit{Step 3 Rebin and combine the spectra.} The above data reduction steps were performed separately on the two transit datasets. We combined the spectra by binning the two transit datasets with a orbital phase step of 0.01. The spectra within each phase bin were added up with their SNRs as weights. In this way, we obtained 18 spectra from phase -0.08 to phase +0.10.There is no data for the phase bin at -0.01. %
These spectra are finally displayed in Fig.1a.

Supplementary Fig.1 shows an example of the data reduction procedures.

\subsection{Stellar spectral model.} 
We simulated the variation of stellar lines ($S_v$) during transit using the method described in \cite{Yan2017} and \cite{Casasayas-Barris2017}. The model includes RM effect and CLV effect. The simulation includes the following steps.

(1) We firstly obtained the stellar spectrum as a function of limb angle using the Spectroscopy Made Easy tool\cite{Piskunov2016}. The simulation utilizes the VALD line list\cite{Ryabchikova2015} and the Kurucz ATLAS code\cite{Kurucz2005}. The stellar parameters used are $T_\mathrm{eff} = 10 170$ K, [Fe/H] = -0.03, log $g$ = 4.093. The stellar rotation was not included in this step, but will be included in the next step.  We simulated the stellar spectra at 21 limb angles and linearly interpolated the spectra at other limb angles. 

(2) We then calculated the synthetic spectrum during transit. We divided the stellar disk into elements with a size of 0.01 $R_\mathrm{star}$ $\times$ 0.01 $R_\mathrm{star}$. The stellar spectrum during transit was then calculated by integrating the spectra of the unobscured elements. The RM effect was included by assigning each stellar element a RV due to the stellar rotation. The parameters used are: $R_\mathrm{star} = 2.362 ~ R_\mathrm{Sun}$, $R_\mathrm{planet} = 1.891 ~ R_\mathrm{Jupiter}$, $i$ = 86.79 degrees, $v~\mathrm{sin}I = 111.4$ km/s and $\lambda$ = -84.8 degrees. 

(3) We obtained the normalized stellar line profile change by dividing each simulated spectrum with the out-of-transit spectrum. The simulated result is presented in Supplementary Fig.2. The RM effect causes a bright Doppler trace. The green line indicates the RV of the rotating stellar surface that is blocked by the planet. This RV value changes from -29 km/s to -8 km/s during transit. The CLV effect makes the line profile shallower at mid-transit while deeper when transiting the stellar limb.

The above simulated stellar line profile change is for 1 $R_\mathrm{planet}$. However, the actual stellar $\mathrm{H_\alpha}$ line change should be larger than the 1 $R_\mathrm{planet}$ case because the planetary atmosphere absorption increases the effective radius. 
In order to properly account for this ``effective radius increase'' problem, we introduced a factor $f$ and assumed the corresponding line profile change is $f$ times of the simulated result for the 1 $R_\mathrm{planet}$ case.
We note here the actual effective radius is quite complicated because one needs to consider the line profile of the planetary absorption (i.~e. the effective radius varies from the line centre to the line wing).  The RV difference between the planetary atmosphere and the stellar elements obscured by the planet (i.~e. the two dashed lines in Fig.1b) also affects the actual effective radius. Therefore, we used a simple constant $f$ factor as a first-order approximation to account for this problem.

\subsection{MCMC analysis.} 
We performed an MCMC analysis to fit the observed data $R_v$ (Fig.~1a) and retrieve the errors. The model includes two parts: the planetary transmission spectrum $T_v$ and the stellar line change $S_v$. The transmission spectrum is assumed to be a Gaussian function:
\begin{equation}
      T_v = 1 + h~e^{-\frac{(v-v_\mathrm{0})^2}{ 2\sigma} }
   \end{equation}
where $v$ is velocity, $\sigma$ is the Gaussian standard deviation, $h$ is the Gaussian height and $v_\mathrm{0}$ is the planetary radial velocity (the planetary orbital motion plus the wind speed).
The final model is obtained by multiplying  the absorption spectrum ($T_v$) and the simulated stellar line variation ($S_v$)

In total, there are five free parameters in the model:
Gaussian standard deviation ($\sigma$), Gaussian height ($h$), planetary orbital motion amplitude ($K_\mathrm{planet}$), wind speed ($v_\mathrm{wind}$) and the stellar model factor ($f$). We only used the fully in-transit data, because the planetary absorption spectrum during ingress and egress is different from the absorption spectrum during fully in-transit. This means the analysis is based on binned spectra covering the phases from -0.045 to +0.045.

The \textit{emcee} code was employed to run the MCMC simulations. 
We used 100 walkers and each walker has 10000 steps. The first 1000 steps were discarded as burn-in. The correlation diagrams are presented in Supplementary Fig.3.

\subsection{Equivalent width calculation.} 
The $\mathrm{H_\alpha}$ equivalent width for each absorption spectrum is calculated by the following equation\cite{Cauley2017}:
\begin{equation}
      W_\mathrm{H_\alpha} = \sum_{v=-75 + v_0}^{+75 + v_0} (1 - R_v^\prime) \Delta \lambda_v
   \end{equation}
where  $R_v^\prime$ is the observed absorption spectrum at velocity $v$, $\Delta \lambda_v$ is the wavelength difference at $v$, $v_0$ is the planetary velocity (combination of orbital motion and wind speed). Here, $R_v^\prime$ is corrected with the modeled stellar line profile change, i.~e. $R_v^\prime=(F_v/F_v^\mathrm{out})/S_v$. The integration range was chosen to be $\pm 75$ km/s considering the actual absorption profile presented in Fig.2, i.~e. this range includes the majority of the line profile. The $W_\mathrm{H_\alpha}$ has a value in the same unit as $\Delta \lambda_v$. We derived the uncertainties of  $W_\mathrm{H_\alpha}$ by summing  $E(R_v) \times  \Delta \lambda_v$ values in quadrature (i.~e. square root of sum of squares), where $E(R_v)$ is the uncertainty of $R_v$. The result is presented in Fig.3. 

\subsection{Results of individual night.} 
The results of individual night are presented in Supplementary Fig.4.
The spectra here are binned with 0.01 phase steps. The transmission spectra and equivalent widths are calculated after the correction of the stellar line variation and planetary orbital motion. We adopted the values of the planetary orbital motion amplitude ($K_\mathrm{planet}$), the wind speed ($v_\mathrm{wind}$) and the stellar model factor ($f$) which are obtained from the combined two nights spectra.
The $\mathrm{H_\alpha}$ signal can be seen clearly in both nights with a similar absorption depth, only that the Night 2 result is more noisy due the cloud passing by during the observation.

The bottom panel of Supplementary Fig.4 presents the equivalent widths of un-binned spectra. As can be seen, the observations are not consecutive due to the interruption of cloud and the spectral SNR has a large variation.

\subsection{Code availability.}
The spectral reduction was done with a self-written IDL script. 
The MCMC analysis was done with the \textit{emcee} tool which is public available from https://github.com/dfm/emcee.
The stellar model spectrum was obtained with the  Spectroscopy Made Easy tool which is public available from
http://www.stsci.edu/$\sim$valenti/sme.html.

\subsection{Data availability.}
The data that support the plots within this paper and other findings of this study are available from the corresponding author upon reasonable request.

\end{methods}

\pagebreak

\begin{figure}
\includegraphics[height=0.65\textwidth, width=0.90\textwidth]{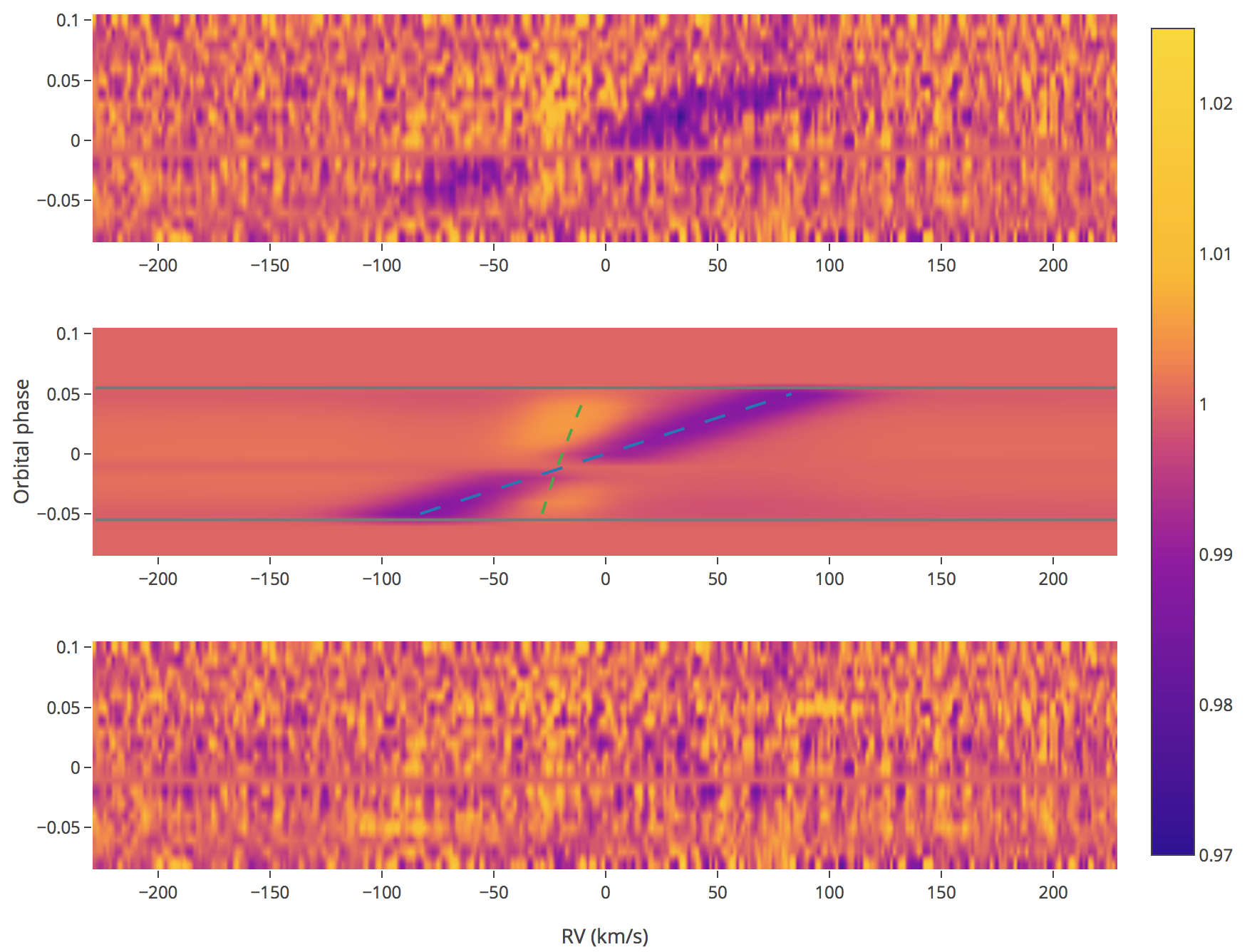} 
\caption{\textbf{The $\mathrm{H_\alpha}$ absorption spectra of KELT-9b at different orbital phases.}
(a) Upper panel: The observed result. These are the ratios ($R_v$) between the observed spectra and the reference out-of-transit spectrum. We binned the data into 0.01 phase steps. There is no data for the phase bin at -0.01. The x axis is wavelength expressed in RV relatively to the $\mathrm{H_\alpha}$ line centre (6562.81 $\mathrm{\AA}$).
(b) Middle panel: The best-fit model from MCMC analysis. The model includes the $\mathrm{H_\alpha}$ transmission spectrum ($T_v$) and the stellar line profile change ($S_v$).  The stellar line profile change is due to the RM effect and the CLV effect. 
The two horizontal lines indicate the beginning and end of transit.
The blue dashed line indicates the RV of the planetary orbital motion plus a constant wind speed.
The green dashed line indicates the RV of the rotating stellar surface which is obscured by the planet. The RV difference explicitly separates the planetary absorption from the stellar line profile change. 
(c) Lower panel: The residual between the observation and the model.
}
\end{figure}

\begin{figure}
\includegraphics[height=0.50\textwidth, width=0.80\textwidth]{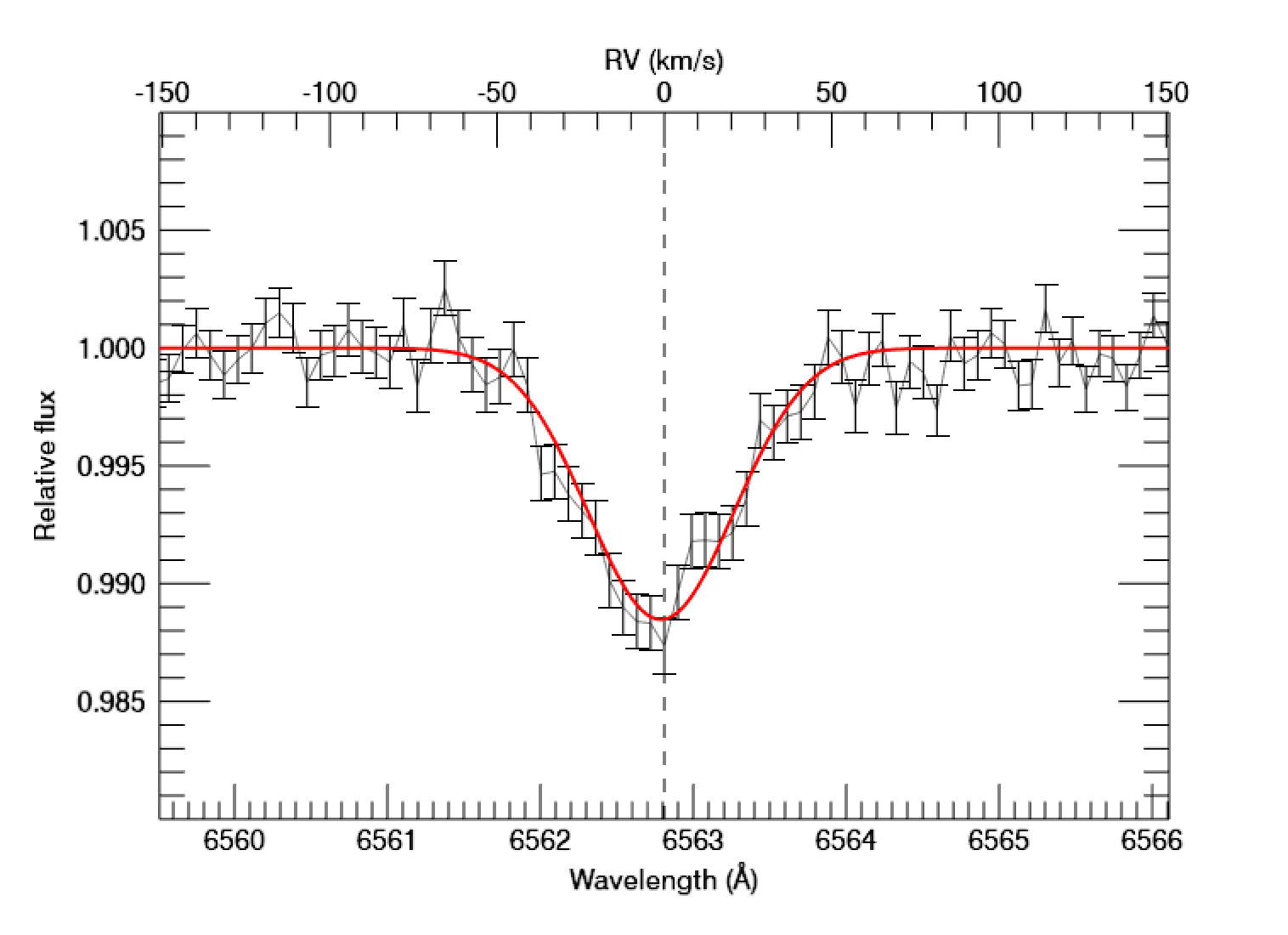} 
\caption{\textbf{The combined $\mathrm{H_\alpha}$ absorption spectrum.}
The spectrum is obtained by adding up all the fully in-transit spectra (8 binned spectra in total). The planetary orbital motion RV was corrected before adding up. We rebinned every three data points. The spectrum has a full width half maximum of $51.2_{-2.5}^{+2.7}~\mathrm{km/s}$ and a line depth of $1.15 \pm 0.05\%$.
The red line is the Gaussian function from the MCMC analysis. 
The vertical dashed line indicates the theoretical $\mathrm{H_\alpha}$ line centre. The observed line centre has an offset of $-1.02_{-1.00}^{+0.99}~\mathrm{km/s}$ compared to the theoretical line centre. 
}
\end{figure}

\begin{figure}
\includegraphics[height=0.50\textwidth, width=0.80\textwidth]{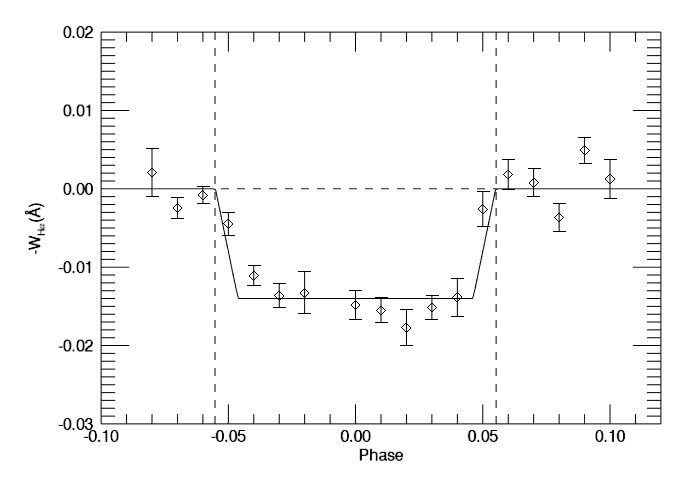} 
\caption{\textbf{The time-series equivalent width of  the $\mathrm{H_\alpha}$ absorption. }
The equivalent width was measured by integrating the absorption feature from -75 km/s to +75 km/s. The modeled stellar line change has been corrected. The two vertical dashed lines indicate the begin and end of  transit. The $W_\mathrm{H_\alpha}$ values are around zero during out-of-transit and are positive during in-transit (note that the ~~ -- $W_\mathrm{H_\alpha}$ is plotted ). The solid line is a transit light curve model with a depth of 0.014  $\mathrm{\AA}$. The light curve is for illustration purpose and limb darkening is ignored.
}
\end{figure}

\begin{figure}
\includegraphics[height=0.50\textwidth, width=0.80\textwidth]{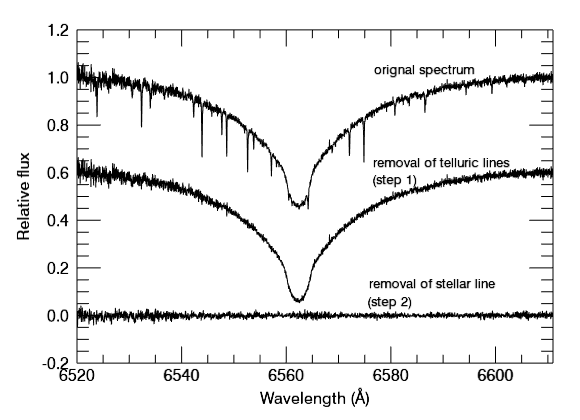} 

\textbf{Supplementary Figure 1} Example of data reduction procedures.
The upper line is an original observed spectrum during out-of-transit. The middle line shows the spectrum after the removal of telluric  $\mathrm{H_2O}$ lines (shifted down by 0.4 for clarity). The bottom line shows the spectrum after the removal of the stellar lines (shifted down by 1.0).
\end{figure}

\begin{figure}
\includegraphics[width=0.90\textwidth] {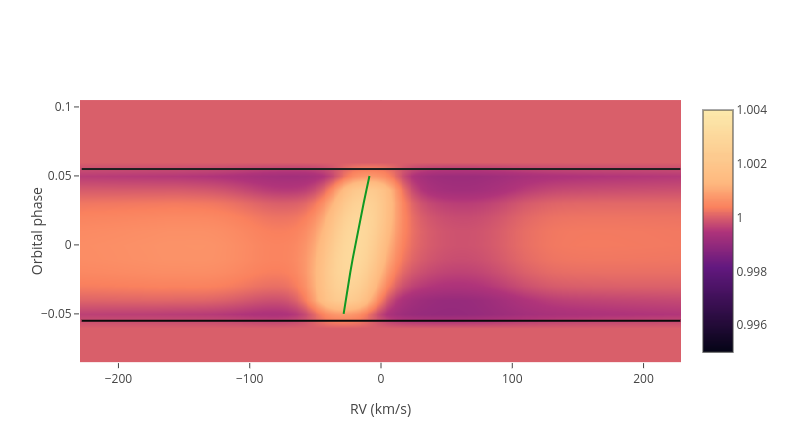} 

\textbf{Supplementary Figure 2} The simulated stellar line profile change during transit.
The two horizontal black lines indicate the beginning and end of transit. The green line indicates the RV of the rotating stellar surface that is obscured by the planet. The simulation is for the 1 $R_\mathrm{planet}$ case.
\end{figure}

\begin{figure}
\includegraphics[width=0.90\textwidth] {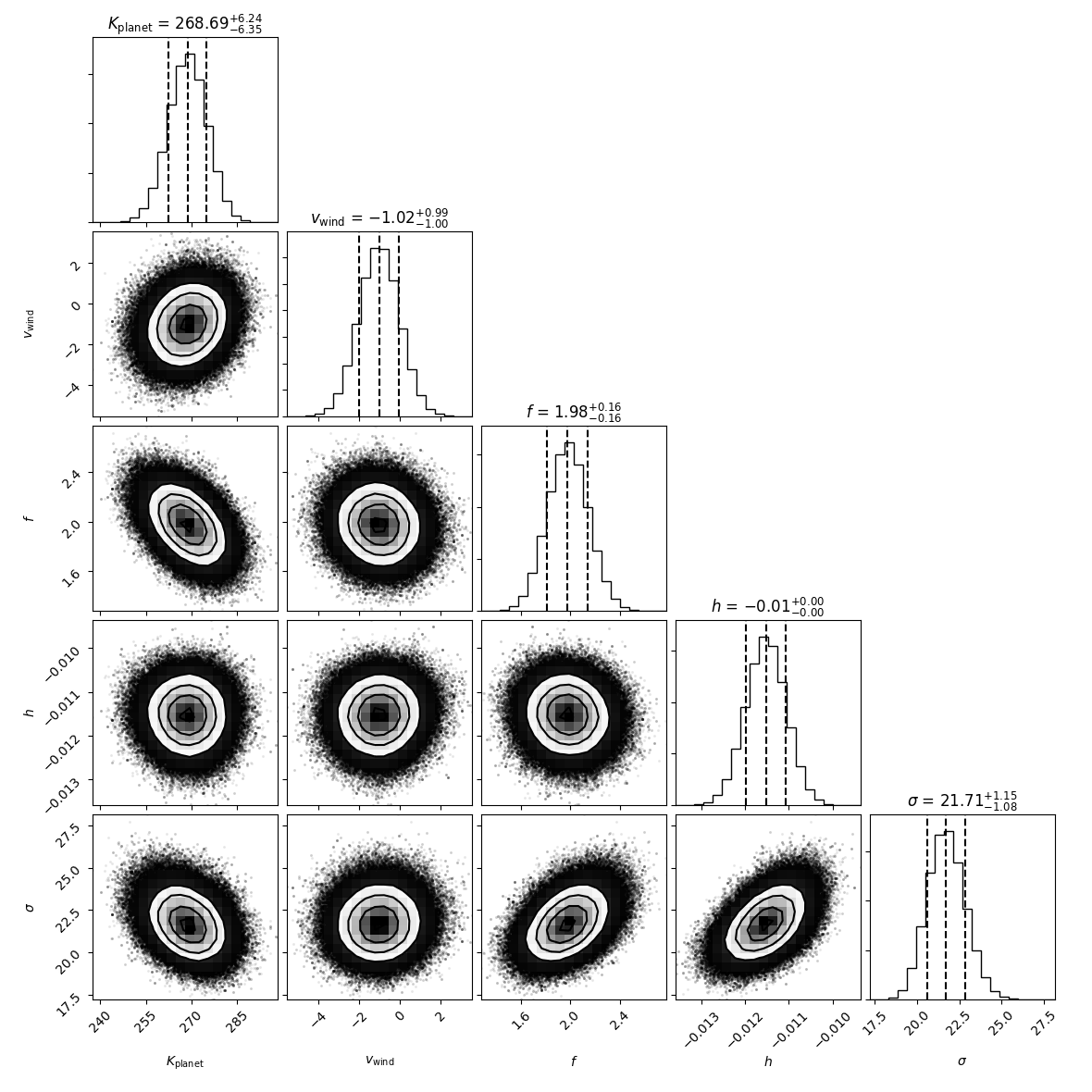} 

\textbf{Supplementary Figure 3} Correlation diagrams from the MCMC analysis.
This is the corner plot of the posterior probability distribution of the five parameters used in the model. The 1 $\sigma$ errors are also indicated as dashed lines.
\end{figure}

\begin{figure}
\includegraphics[width=0.90\textwidth, height=0.95\textwidth] {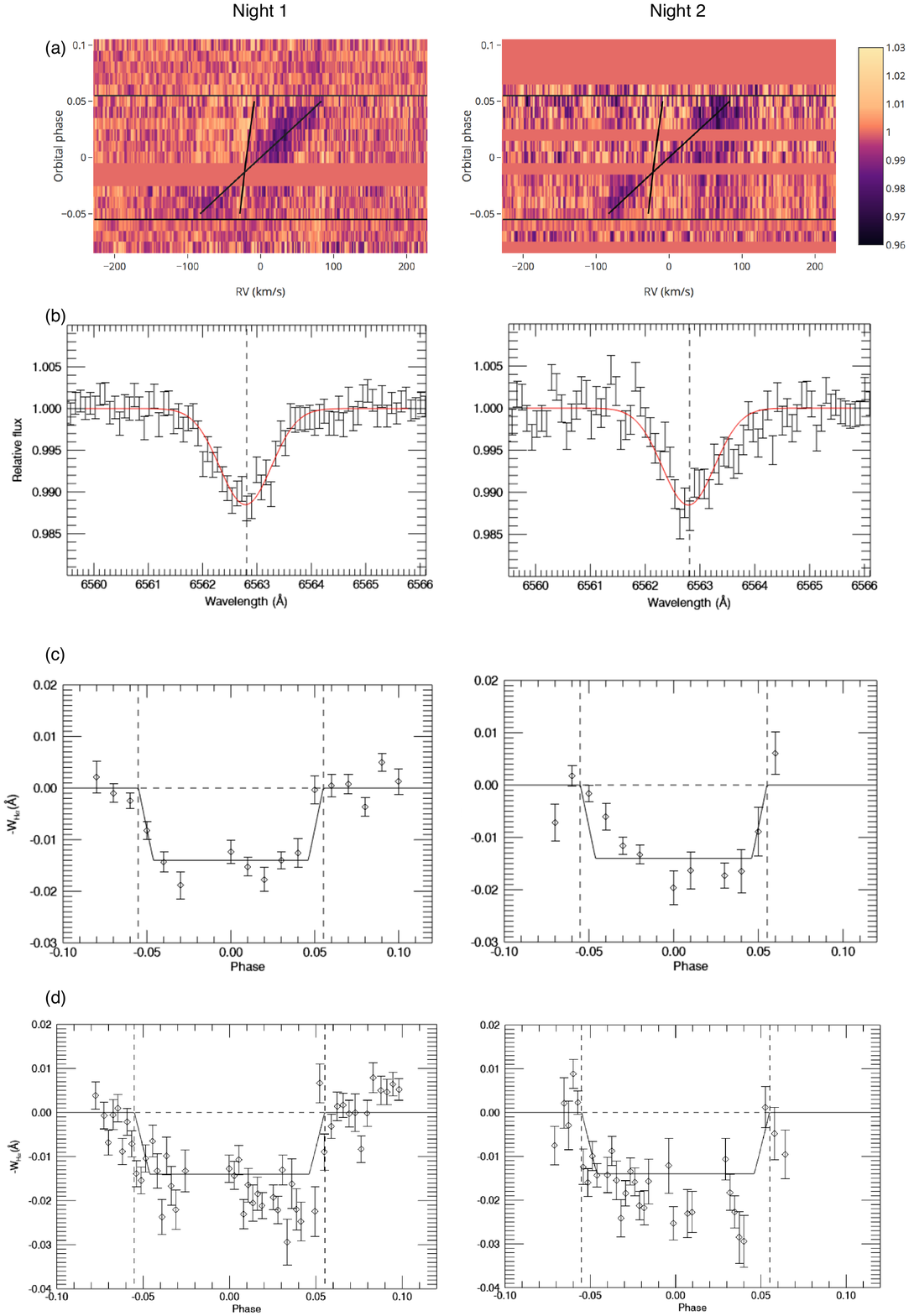} 

\textbf{Supplementary Figure 4} The results of individual nights. The left column is for Night 1 (August 6, 2017) and the right column is for Night 2 (September 21, 2017). (a) The observed spectra binned with 0.01 phase steps. The lines are the same as in Figure 1.
(b) The combined $\mathrm{H_\alpha}$ absorption spectrum.
(c) The time-series equivalent width of  the $\mathrm{H_\alpha}$ absorption.
(d) The time-series equivalent width of  the $\mathrm{H_\alpha}$ obtained from un-binned spectra.
\end{figure}




\end{document}